\documentclass[10pt]{iopart}
\usepackage{epsf,epsfig}

\renewcommand{\d}{{\rm d}}

\newcommand{\1}{{\bf 1}}
\newcommand{\ii}{{\bf i}}
\newcommand{\jj}{{\bf j}}
\newcommand{\kk}{{\bf k}}
\newcommand{\qq}{{\bf q}}
\newcommand{\xx}{{\bf x}}

\begin{document}

\title{Detecting Topology in a Nearly Flat Spherical Universe}

\author{Jeffrey Weeks$^{1}$,
        Roland Lehoucq$^{2,3}$,
        Jean-Philippe Uzan$^{4,5}$
        }

\date{18 September 2002}

\address{(1) 15   Farmer   St.,  Canton  NY  13617-1120  (USA).\\
         (2) CE-Saclay, DSM/DAPNIA/Service d'Astrophysique,
             F-91191 Gif sur Yvette Cedex (France)\\
         (3) Laboratoire Univers et Th\'eories, CNRS-FRE 2462,
             Observatoire  de  Paris, F-92195 Meudon Cedex
             (France)\\
         (4) Institut d'Astrophysique de Paris, GReCO, CNRS-FRE 2435,\\
             98 bis, Bd Arago, 75014 Paris
             (France).\\
         (5) Laboratoire de Physique Th\'eorique, CNRS-UMR 8627,
             B\^at. 210, Universit\'e Paris XI, F--91405 Orsay Cedex
             (France).
        }

\begin{abstract}
When the density parameter is close to unity, the universe has a
large curvature radius independently of its being hyperbolic,
flat, or spherical. Whatever the curvature, the universe may have
either a simply connected or a multiply connected topology. In the
flat case, the topology scale is arbitrary, and there is no {\it a
priori} reason for this scale to be of the same order as the size
of the observable universe. In the hyperbolic case any nontrivial
topology would almost surely be on a length scale too large to
detect. In the spherical case, by contrast, the topology could
easily occur on a detectable scale. The present paper shows how,
in the spherical case, the assumption of a nearly flat universe
simplifies the algorithms for detecting a multiply connected
topology, but also reduces the amount of topology that can be
seen. This is of primary importance for the upcoming cosmic
microwave background data analysis.

This article shows that for spherical spaces one may restrict the
search to diametrically opposite pairs of circles in the
circles-in-the-sky method and still detect the cyclic factor in
the standard factorization of the holonomy group. This vastly
decreases the algorithm's run time.  If the search is widened to
include pairs of candidate circles whose centers are almost
opposite and whose relative twist varies slightly, then the cyclic
factor along with a cyclic subgroup of the general factor may also
be detected. Unfortunately the full holonomy group is, in general,
unobservable in a nearly flat spherical universe, and so a full
6-parameter search is unnecessary. Crystallographic methods could
also potentially detect the cyclic factor and a cyclic subgroup of
the general factor, but nothing else.
\end{abstract}


\section{Introduction}

Recent cosmic microwave background (CMB) data analysis suggests an
approximately flat universe. The constraint on the total density
parameter, $\Omega$\footnote{$\Omega$ is the ratio between the
total energy density and the critical energy density of the
universe.}, obtained from CMB experiments depends on the priors
used during the data analysis. For example with a prior on the
Hubble parameter and on the age of the universe, recent
analysis~\cite{sievers,netterfield} of the DASI, BOOMERanG, MAXIMA
and DMR data lead to $\Omega=0.99\pm0.12$ at 1$\sigma$ level and
to $\Omega=1.04\pm0.05$ at 1$\sigma$ level if one takes into
account only the DASI, BOOMERanG and CBI data. Including stronger
priors can indeed sharpen the bound. For instance, including
information respectively on large scale structure and on
supernovae data leads to $\Omega=1.01_{-0.06}^{+0.09}$ and
$\Omega=1.02_{-0.08}^{+0.09}$ at 1$\sigma$ level while including
both finally leads to $\Omega=1.00_{-0.06}^{+0.10}$. In
conclusion, it is fair to retain that current cosmological
observations only set the bound $0.9 <\Omega< 1.1$. These results
are consistent with Friedmann-Lema\^{\i}tre universe models with
spherical, flat or hyperbolic spatial sections.  In the spherical
and hyperbolic cases, $\Omega \approx 1$ implies that the
curvature radius must be larger than the horizon radius. In all
three cases -- spherical, flat, and hyperbolic -- the universe may
be simply connected or multiply connected, but our chances of
detecting the topology observationally depend strongly on the
curvature.

The chances of detecting a multiply connected topology are worst
in a large hyperbolic universe.  The reason is that the typical
translation distance between a cosmic source and its nearest
topological image seems to be on the order of the curvature
radius, but when $\Omega \approx 1$ the horizon radius is less
than half that distance.  For example, if $\Omega_m = 0.34$,
$\Omega_\Lambda = 0.64$ and $\Omega = 0.98$, then the
radius $\chi_{\rm LSS}$ of the last scattering surface is 0.43
radians \footnote{A hyperbolic radian is defined to equal the
curvature radius, just like a spherical radian.} (see
Eq.~\ref{chiLSS} below).  As $\Omega$ approaches 1,
$\chi_{\rm LSS}$ approaches 0 radians, which is much less than the
typical topology scale, making the topology undetectable (see e.g.
Refs~\cite{gomero1,aurichsteiner,inoue3} for some studies on
detectability of nearly flat hyperbolic universes). Note that, in
spite of a widespread myth to the contrary, the topology scale in
a hyperbolic 3-manifold might {\it not} depend on the manifold's
volume. It might remain comparable to the curvature radius even
for large manifolds, assuming the observer sits at a generic point
in the manifold. On the other hand, in the exceptional case that
the observer sits near a short closed geodesic, his or her nearest
translated image may be arbitrarily close, even in an arbitrarily
large manifold.

In a multiply connected flat universe the topology scale is completely
independent of the horizon radius, because Euclidean geometry --
unlike spherical and hyperbolic geometry -- has no preferred scale and
admits similarities. A great deal of luck would be required for the
topology scale to be less than the horizon radius but still large
enough to accommodate the lack of obvious local
periodicity\footnote{The constraint on the size of a cubic 3-torus
varies from half of the horizon size~\cite{tore1} to about one
fourth~\cite{tore2} depending on the value of the cosmological
constant. In the case of a vanishing cosmological constant, these
constraints were extended to other flat topologies~\cite{flat}}.
Detecting such a topological structure would naturally raise deep
questions about this ``topological" coincidence: why is the topology
scale of the order of the size of the observable universe today?

In a spherical universe the topology scale depends on the
curvature radius, as in the hyperbolic case. Luckily, in contrast
to the hyperbolic case, as the topology of a spherical manifold
gets more complicated, the typical distance between two images of
a single cosmic source decreases. No matter how close $\Omega$ is
to 1, only a finite number of spherical topologies are excluded
from detection. For example, if $\Omega_m = 0.34$, $\Omega_\Lambda
= 0.68$ and $\Omega = 1.02$, then $\chi_{\rm LSS}$ =
0.43 radians, and the only excluded topologies are those for which
the cyclic factor (see Section~\ref{SectionClassification} for
explanations) has order at most ${2\pi}/({2 \chi_{LSS}}) \approx
7$. As $\Omega$ gets closer to 1, more topologies are
excluded from observation; nevertheless the chances of observing a
spherical universe remain vastly better than the chances of
observing a flat universe, because in the flat case only a finite
range of length scales are short enough to be observable while an
infinite range is too large. The particular case of the
detectability of lens spaces was studied in Ref.~\cite{gomero2}
(which also considers the detectability of hyperbolic topologies).

The present paper investigates how the assumption of a nearly flat
($\Omega \approx 1)$ spherical universe affects the strategy for
detecting its topology (see Refs.~\cite{texas,luminet,levin} for
reviews on the different methods and their observational status).
We find that the circles-in-the-sky
method~\cite{CornishSpergelStarkman} typically detects only a
cyclic subgroup of the holonomy group, so the universe ``looks
like a lens space'' no matter what its true topology is. If it
looks like a globally homogeneous lens space (generated by a
Clifford translation -- to be defined in Section
\ref{SectionClifford}), then the circles-in-the-sky search reduces
from a 6-parameter search space to a 3-parameter space (if the
radius $\chi_{\rm LSS}$ of the last scattering surface is known)
or a 4-parameter space (if $\chi_{\rm LSS}$ is unknown). Assuming
each parameter is tested at $\sim 10^3$ points, the total run time
decreases by a factor of $\sim 10^9$ or $\sim 10^6$. If the
universe looks like a non globally homogeneous lens space, meaning
that a larger subgroup of the holonomy group is detectable,
generated by a pair of Clifford translations, then the matching
circles lie in a 6-parameter space, but with strong constraints on
their values, so the search is still much faster than in the
general case.

We also note that the crystallographic method's pair separation
histogram~\cite{LLL} is well suited to detect a topological signal
in a nearly flat spherical universe because of the abundance of
Clifford translations (defined below).

Section~\ref{SectionClassification} reviews the classification of
spherical 3-manifolds. Section~\ref{SectionClifford} defines
Clifford translations and explains their cosmological
significance. Section~\ref{SectionCircles} analyzes the
circles-in-the-sky method, and Section
\ref{SectionCrystallography} analyzes the crystallographic
methods.

\section{Classification of Spherical 3-Manifolds}
\label{SectionClassification}

Spherical 3-manifolds were originally classified by Threlfall and
Seifert \cite{ThrelfallSeifert} in 1930.  Gausmann {\it et al.}
\cite{GLLUW} reformulates Threlfall and Seifert's classification in
terms of single action, double action, and linked action
manifolds. This latter work also describes the manifolds in each category,
and investigates the topological signature of each in the pair
separation histogram of cosmic crystallography.

Interactive software for visualizing these manifolds may be
downloaded at \texttt{www.northnet.org/weeks/CurvedSpaces}.  The
user flies around in spherical 3-manifolds to gain intuition about
their topological properties. For instance, the role of a cyclic
factor (as described below) can be well understood with this tool;
screenshots appear in
Figs.~(\ref{FigureStrings}-\ref{FigureSaddleNbr}).

In this section, we briefly review the classification given as given
in Ref.~\cite{GLLUW}.  Even though we could follow Threlfall and
Seifert's purely geometric approach, i.e. working with right- and
left-handed screw motions, we instead borrow from Thurston's approach
\cite{Thurston} and use quaternions to streamline the exposition.

Recall that quaternions are like complex numbers, except that
while the complex numbers are spanned by two elements \1 and \ii\,
satisfying $\ii^2 = -1$, the quaternions are spanned by four
elements elements \1, \ii, \jj, and \kk\, satisfying the
(non-commutative) multiplication rules
\begin{eqnarray}\label{eq:quaternion1}
&\ii^2 = \jj^2 = \kk^2 = -1& \nonumber\\
&\ii\jj = -\jj\ii = \kk& \nonumber\\
&\jj\kk = -\kk\jj = \ii& \nonumber\\
&\kk\ii = -\ii\kk = \jj&
\end{eqnarray}
A quaternion $\qq = a\1 + b\ii + c\jj + d\kk$, with $a,b,c,d$
being real numbers, has length $|\qq| = \sqrt{a^2 + b^2 + c^2 +
d^2}$. The set of unit length quaternions defines the Lie group
${\cal S}^3$, combining the geometry of the 3-sphere $S^3$ with
the multiplicative action of the quaternions, just as the set of
unit length complex numbers defines the Lie group ${\cal S}^1$,
combining the geometry of the circle $S^1$ with the multiplicative
action of the complex numbers.

\subsection{Single action manifolds}
\label{SubsectionSingleAction}

Each quaternion, $\qq \in {\cal S}^3$, defines one isometry,
$\ell_\qq: {\cal S}^3 \rightarrow {\cal S}^3$, by left multiplication
$\ell_\qq(\xx) = \qq\xx$
\begin{equation}
\ell_\qq:
\left\lbrace
\begin{array}{l}
{\cal S}^3 \rightarrow {\cal S}^3\\
\xx\mapsto\ell_\qq(\xx) = \qq\xx.
\end{array}
\right.
\end{equation}
Another isometry, $r_\qq: {\cal S}^3 \rightarrow {\cal S}^3$, can
analogously be defined by right multiplication $r_\qq(\xx) =
\xx\qq$. If $G$ is a finite subgroup of ${\cal S}^3$, then $\Gamma =
\{\ell_\qq\,|\,\qq \in G\}$ is a finite group of isometries of $S^3$
(and isomorphic to ${\cal S}^3$). Except for the identity, the
isometries in $\Gamma$ have no fixed points, so that $\Gamma$ defines
the holonomy group\footnote{It is also referred to as group of
covering transformations.} of a spherical 3-manifold. The manifolds
that arise in this way are called {\it single action} spherical
3-manifolds. Had we used $r_\qq$ instead of $\ell_\qq$ in the
definition of $\Gamma$ we would have obtained the mirror image of the
same manifold.

The single action manifolds are in one-to-one correspondence with the
finite subgroups of ${\cal S}^3$, listed in Table
\ref{TableS3Subgroups}.  Section 3 of Ref.~\cite{GLLUW} explains the
relationship between these subgroups of Isom($S^3$) and the
corresponding subgroups of Isom($S^2$), while Appendix B of
Ref.~\cite{GLLUW} explicitly determines the elements in each group
both as quaternions and as $SO(4)$-matrices.

\begin{table}
\begin{center}
\begin{tabular}{|l|c|c|c|}
\hline
Name  of the manifold   & Symbol  & Order & Fundamental domain \\
\hline
cyclic                  & $Z_n$   & $n$  & lens             \\
binary dihedral         & $D^*_m$ & $4m$ & $2m$-sided prism \\
binary tetrahedral      & $T^*$   & 24   & octahedron       \\
binary octahedral       & $O^*$   & 48   & truncated cube   \\
binary icosahedral      & $I^*$   & 120  & dodecahedron     \\
\hline
\end{tabular}
\end{center}
\caption{Finite subgroups of ${\cal S}^3$. There is a two-to-one
homomorphism from ${\cal S}^3$ to $SO(3)$ (see Ref.~\cite{GLLUW}
for details) so the finite subgroups of $SO(3)$ -- the cyclic,
dihedral, tetrahedral, octahedral and icosahedral groups -- lift
to ${\cal S}^3$.  Each such group lifts two-to-one, giving the
corresponding binary group. Only the cyclic groups of odd order
lift one-to-one. That is, each even-order cyclic group $Z_{2n}
\subset {\cal S}^3$ arises as the two-fold cover of $Z_n \subset
SO(3)$, while each odd-order cyclic group $Z_n \subset {\cal S}^3$
is the one-to-one lift of $Z_n \subset SO(3)$.  The binary
polyhedral groups are not merely the product of the original
polyhedral groups with a $Z_2$ factor, nor are they isomorphic to
the so-called extended polyhedral groups containing
orientation-reversing as well as orientation-preserving
isometries; rather they are something completely new. The plain
dihedral, tetrahedral, octahedral and icosahedral groups do not
occur as subgroups of ${\cal S}^3$.} \label{TableS3Subgroups}
\end{table}

\subsection{Double action manifolds}
\label{SubsectionDoubleAction}

In a {\it double action} spherical 3-manifold two groups act
simultaneously, one by left multiplication and the other by right
multiplication.  That is, each pair of quaternions $(\qq, \qq ')
\in G \times G' \subset {\cal S}^3 \times {\cal S}^3$ defines an
isometry by simultaneous left and right multiplication
\begin{equation}
d_{\qq,\qq '}:
\left\lbrace
\begin{array}{l}
{\cal S}^3 \rightarrow {\cal S}^3\\
\xx\mapsto d_{\qq,\qq '}(\xx) = \qq\xx\qq'.
\end{array}
\right.
\end{equation}
The resulting group $\Gamma = \{d_{\qq,\qq '}\,|\,(\qq, \qq ') \in G
\times G'\}$ will be fixed point free (and thus define a manifold) if
and only if the orders of $G$ and $G'$ are relatively prime, with the
possible exception of a common factor of two corresponding to the
quaternion $-\1$. The binary polyhedral groups all contain elements of
order~4 (see Table~\ref{TableS3Subgroups}) and therefore cannot be
paired with each other.  Thus either $G$ or $G'$ must be cyclic.
Without loss of generality we henceforth assume that in a double
action manifold the second factor $G'$ is cyclic. The first factor $G$
may be either cyclic or binary polyhedral.

\subsection{Linked action manifolds}
\label{SubsectionLinkedAction}

{\it Linked action} spherical 3-manifolds are similar to double
action manifolds.  One begins with a pair of groups $G$ and $G'$
whose orders are {\it not} relatively prime, but then takes
$\Gamma \subset \{d_{\qq,\qq '}\,|\,(\qq, \qq ') \in G \times G'\}$
to be a subgroup avoiding elements $d_{\qq,\qq '}$ with fixed points.
The only examples of linked action manifolds are the following:
\begin{itemize}
\item
    $G$ and $G'$ are both cyclic and $\Gamma$ is the holonomy group
    of a lens space $L(p,q)$.
\item
    $G = T^*$, $G' = Z_{9n}$ ($n$ odd), and $\Gamma$ is a
    (fixed point free) subgroup of index 3 in the full
    (fixed point containing) double action group.
\item
    $G = D^*_m$, $G' = Z_{8n}$ ($m$ odd and relatively prime
    to $n$), and $\Gamma$ is a
    (fixed point free) subgroup of index 2 in the full
    (fixed point containing) double action group.
\end{itemize}
Section 4.3 of Ref.~\cite{GLLUW} explains these manifolds in greater
detail.

\section{Clifford Translations}
\label{SectionClifford}

Each left multiplication $\ell_\qq$ defines a {\it Clifford
translation} because it translates all points the same distance --
that is, ${\rm dist}(\xx,\ell_\qq(\xx))$ is constant for all $\xx
\in {\cal S}^3$ -- and similarly for each right multiplication
$r_\qq$. Clifford translations play a crucial role in cosmic
topology. As will be detailed later, the circles-in-the-sky
method~\cite{CornishSpergelStarkman} works best with them, and the
pair separation histogram (PSH) method of cosmic crystallography
detects only Clifford translations~\cite{lehoucq99}.

\begin{figure}
\begin{center}
 \epsfig{file=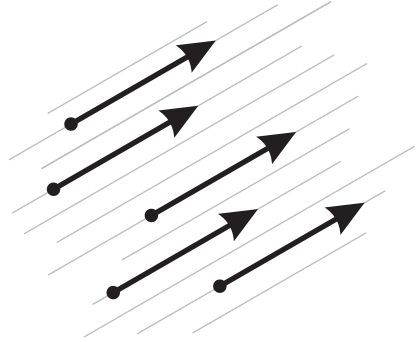} \caption{A translation of the Euclidean
 plane $E^2$ (characterized by a vector $\vec u$) defines a family of
 parallel flow lines, in the sense that to any point $M_0\in E^2$ one
 can associate the flow $M(t)=M_0+t\vec u$.}
 \label{FigureTranslationE2}
\end{center}
\end{figure}

\begin{figure}
\begin{center}
 \epsfig{file=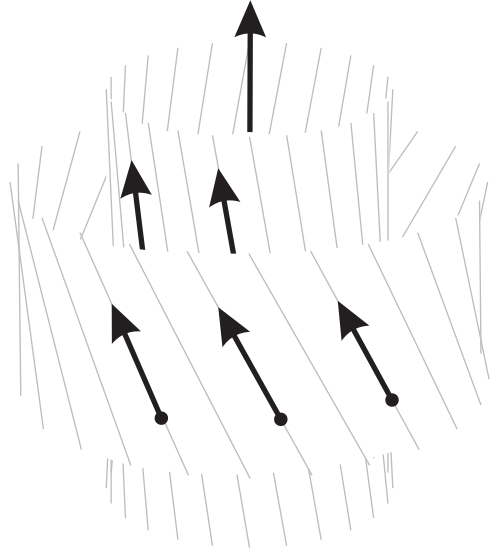} \caption{A Clifford translation of the
 3-sphere $S^3$ defines a family of Clifford parallels. This figure shows
 a left-handed flow of Clifford parallels (i.e. the
 flows are spiralling counterclockwise) corresponding to a right
 multiplication $r_{\qq(t)}$.}  \label{FigureTranslationS3}
\end{center}
\end{figure}

Just as a Euclidean translation defines a family of parallel lines
in the Euclidean plane (Figure~\ref{FigureTranslationE2}), a left
multiplication $\ell_\qq$ of ${\cal S}^3$ defines a family of {\it
Clifford parallels} on the 3-sphere
(Figure~\ref{FigureTranslationS3}). To make this precise, replace
the static quaternion $\qq = a\1 + b\ii + c\jj + d\kk$ with a
time-dependent quaternion
\begin{equation}
\qq(t)=  \cos{2\pi t} \, \1
 + \sin{2\pi t}\frac{b\,\ii +c\,\jj +d\,\kk}{\sqrt{b^2 + c^2 +
 d^2}},
\end{equation}
to generate a flow on $S^3$ carrying each point $\xx\in{\cal S}^3$
along the flow line $\xx(t)=\qq(t)\xx$.  The flow lines comprise a
family of geodesics (i.e. great circles) called {\it Clifford
parallels}. The flow is homogeneous in the sense that any Clifford
parallel is equivalent to any other.  Any two nearby Clifford
parallels twist around each other exactly once.  This remarkable
behavior characterizes the geometry of the 3-sphere,
distinguishing it from familiar Euclidean geometry.

Each left multiplication $\ell_\qq$ defines a right handed family
of Clifford parallels, in the sense that as you move forward along
a given parallel, neighboring parallels twist clockwise.
Similarly, each right multiplication $r_\qq$ defines a left handed
family of Clifford parallels.  Henceforth all discussion of
$\ell_\qq$ will apply equally well to $r_\qq$, but with opposite
chirality.  Note that left multiplication defines right handed
Clifford parallels while right multiplication defines left handed
Clifford parallels. Strictly speaking, an isometry $\ell_\qq$
takes the 3-sphere and simply places it in its new position.  In
practice, though, one visualizes $\ell_\qq$ as a continuous motion
$\ell_{\qq(t)}$ , sliding the 3-sphere from its original position
to its final position along the Clifford parallels. Each point
moves along a geodesic while nearby points spiral around it.
Because of this visual image, each $\ell_\qq$ (resp. $r_\qq$) is
called a {\it right handed} (resp. {\it left handed}) {\it
Clifford translation}.

If $\qq \neq \pm\1$ there is a unique right handed Clifford
translation $\ell_\qq$ (and a unique left handed Clifford
translation $r_\qq$) taking $\1$ to $\qq$, the corresponding
Clifford parallels are well defined, and $\ell_\qq \neq r_\qq$.
The quaternion $-\1$ is exceptional:  both $\ell_{-\1}$ and
$r_{-\1}$ are the antipodal map and correspond to translation
half way along {\it any} set of Clifford parallels.
This degeneracy accounts for an exceptional factor of two that
runs throughout the theory of spherical 3-manifolds (see Ref.~\cite{GLLUW}).
The quaternion $\1$ is also exceptional in that
$\ell_\1 = r_\1 = {\rm Id}$ and corresponds to the null translation
on any set of Clifford parallels, but because $\1$ is already
the identity quaternion it creates no exceptional cases.

\begin{figure}
\begin{center}
\epsfig{file=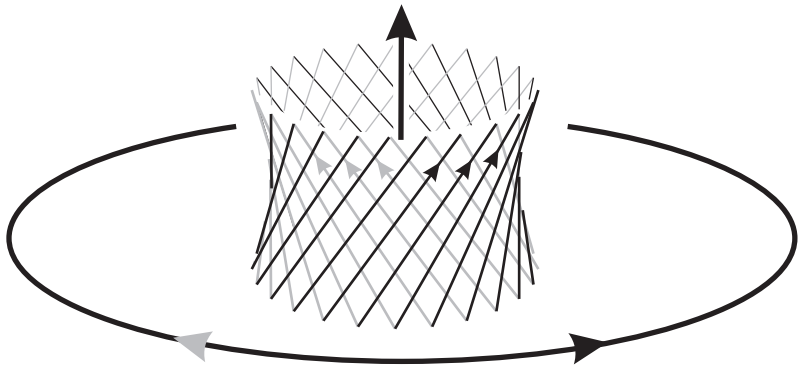} \caption{Each set of right handed
Clifford parallels (black) has exactly two great circles in common
with each set of left handed Clifford parallels (grey).  In the
neighborhood of one common great circle (central axis) the right
handed Clifford parallels twist clockwise while the left handed
Clifford parallels twist counterclockwise.  In the neighborhood of
the other common great circle (equatorial belt) the right and left
handed Clifford parallels still twist clockwise and
counterclockwise, respectively, but translate in opposite
directions.} \label{FigureParallels}
\end{center}
\end{figure}

Any set of right handed Clifford parallels has exactly two great
circles in common with any set of left handed Clifford parallels
(Figure~\ref{FigureParallels}).  Furthermore the two common
circles are maximally far from each other: the distance from any
point on one of the common circles to any point on the other is
exactly $\pi/2$.  In a neighborhood of a common circle, a Clifford
translation along the right handed parallels twists clockwise
while a Clifford translation along the left handed parallels
twists counterclockwise.

\section{Circle Searching in a Nearly Flat Spherical Universe}
\label{SectionCircles}

The circles-in-the-sky method for detecting cosmic topology relies
on the fact that in a sufficiently small universe the last
scattering surface will ``wrap all the way around the universe''
and intersect itself~\cite{CornishSpergelStarkman} (see also
Ref.~\cite{levin} for a recent review of topology and the CMB).
Each circle of self-intersection will appear as two different
circles in the sky with matching CMB temperatures (modulo various
sources of interference, such as the integrated Sachs-Wolfe
effect, the Doppler effect due to the primordial plasma's motion,
and microwave contamination in the plane of the Milky Way). If one
can detect matching circles, one can deduce the topology of
space~\cite{jeff}.

In its most general form, the circles-in-the-sky method searches
over a 6-dimensional parameter space:  two parameters for the
center of the first candidate circle, two parameters for the
center of the second candidate circle, one parameter for the
circles' angular radius, and one parameter for the relative twist
with which the circles are compared.  To test for a nonorientable
topology one must, in addition, compare each pair of candidate
circles with reversed orientation.  A full 6-parameter search is
computationally expensive, and simplifications must be
found~\cite{Cornish}.

In this section, we want to show that circle searching is much
easier in a nearly flat spherical manifold than in the general
case. To fix the notations, the local geometry of the universe is
described by a Friedmann--Lema\^{\i}tre metric
\begin{equation}\label{METR}
 \d s^2=-\d t^2+a^2(t)\left(\d\chi^2+\sin^2{\chi}\d\omega^2
 \right)
\end{equation}
where $a$ is the scale factor, $t$ the cosmic time and
$\d\omega^2\equiv \d\theta^2+\sin^2{\theta}\d\varphi^2$ the
infinitesimal solid angle. $\chi$ is the (dimensionless) comoving
radial distance in units of the curvature radius of $S^3$, that
is, in radians so that $\chi$ runs from 0 to $\pi$. Integrating
the radial null geodesic equation $\d\chi=\d t/a$ leads  to
\begin{equation}\label{chi2}
 \chi(z)=\int_0^z\frac{\sqrt{\Omega_{m}+\Omega_{\Lambda}-1}\,\d x}
 {\sqrt{\Omega_{\Lambda}+(1-\Omega_{m}-\Omega_{\Lambda})(1+x)^2
 +\Omega_{m}(1+x)^3}}
\end{equation}
where $\Omega_\Lambda$ and $\Omega_m$ are the cosmological
constant and matter density parameters evaluated today. The
comoving coordinate of the last scattering surface is thus given
by
\begin{equation}\label{chiLSS}
 \chi_{\rm LSS}=\chi(z_{\rm LSS})
\end{equation}
with $z_{\rm LSS}\sim1100$.

\subsection{Circle searching in a single action manifold}

In a single action manifold, every holonomy is a Clifford
translation. One visualizes a Clifford translation taking one
image of the last scattering surface to another of its images as a
translation along a geodesic combined with a simultaneous twist
about the same geodesic. Thus every matched circle lies directly
opposite its mate. This immediately reduces the parameter space
from 6 to 4 dimensions, because the center of one candidate circle
completely determines the center of the other. If the universe is
nearly flat, the parameter space reduces further: the radius
$\chi_{\rm LSS}$ of the last scattering surface (in radians) is
small, so any pair of matched circles corresponds to a short
translation distance, and hence to a small twist.

\begin{figure}
\begin{center}
 \epsfig{file=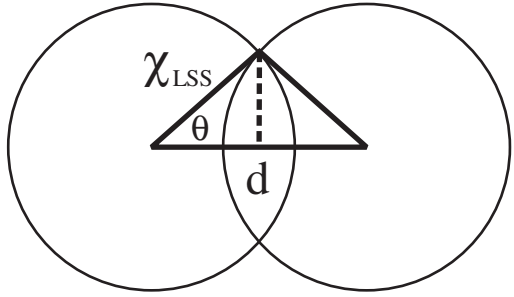}
 \caption{If the radius $\chi_{\rm LSS}$ of the
 last scattering surface is known in advance (in units of the
 curvature radius), then the angular diameter $\theta$ of a matched
 circle determines the distance $d$ to the observer's translated
 image (still in curvature units) via spherical trigonometry:
 $\tan d/2 = \tan \chi_{\rm LSS} \cos\theta$. If the translation is a
 Clifford translation, the twist angle $\varphi$ exactly equals the
 translation distance $d$.}
 \label{FigureTranslationTwistRelation}
\end{center}
\end{figure}

Because Clifford translations translate and twist exactly the same
amount, the twist angle between two matched circles reveals the
translation distance in radians. If the radius $\chi_{\rm LSS}$ of
the last scattering surface (in radians) is known (or well
estimated) ahead of time, then the matched circles' angular radius
completely determines their twist and vice versa (see
Figure~\ref{FigureTranslationTwistRelation}). This further reduces
the search parameter space to 3 dimensions. In other words, the
center of one candidate circle determines the center of the other,
and their common angular radius determines the twist. Furthermore,
the twist is quantized, with values of the form $2\pi/n$ for
integer $n$.

\begin{table}
\begin{center}
\begin{tabular}{|l|c|c|}
\hline
Group & Min. translation distance & Sample min. $\Omega$ \\
\hline
Binary icosahedral $I^*$   & $2\pi/10$ & 1.011 \\
Binary octahedral  $O^*$   & $2\pi/8 $ & 1.017 \\
Binary tetrahedral $T^*$   & $2\pi/6 $ & 1.030 \\
Binary dihedral    $D^*_m$ & $2\pi/2m$ & $\sim (1 + {1}/{(2m)^2})$ \\
Cyclic             $Z_n$   & $2\pi/n$  & $\sim (1 + {1}/{n^2})$ \\
\hline
\end{tabular}
\end{center}
 \caption{For each single action manifold, the second column gives
 the distance from the observer to his or her nearest translated
 image. The third column tells the minimum value of $\Omega$ 
 required for the last scattering surface to intersect
 itself. That is, matched circles will occur in the given topology
 if and only if $\Omega$ exceeds the value given.
 Strictly speaking, the minimum value of $\Omega$
 depends on both $\Omega_m$ and $\Omega_{\Lambda}$; for simplicity,
 the sample values in the third column fix $\Omega_m$ at 0.35 and
 vary $\Omega_{\Lambda}$.}
 \label{TableSingleActionCMB}
\end{table}

Table~\ref{TableSingleActionCMB} shows the distance from a source
to its nearest translate in each of the single action manifolds.
The table's third column shows a typical value of $\Omega$ 
for which the diameter of the last scattering surface
drops below the manifold's minimum translation distance, and the
topology becomes undetectable. If the true $\Omega$
exceeds the minimum then nearby images of the last scattering
surface overlap and the topology is potentially detectable. The
groups $D^*_m$ and $Z_n$ are always detectable for large values of
$m$ and $n$, even though low values of $m$ and $n$ may be
excluded. The minimum values of $\Omega$ shown in the
table were found by using Eq.~(\ref{chiLSS}) to compute the radius
of the last scattering surface in curvature units as a function of
$\Omega_m$ and $\Omega_{\Lambda}$.  Trial and error quickly
revealed which values of $\Omega_m$ and $\Omega_{\Lambda}$ were
required for the diameter of the last scattering surface to equal
the minimum translation distance shown in second column of the
table.

\subsection{Circle searching in a double action or linked action manifold}

\begin{figure}
\begin{center}
 \epsfig{file=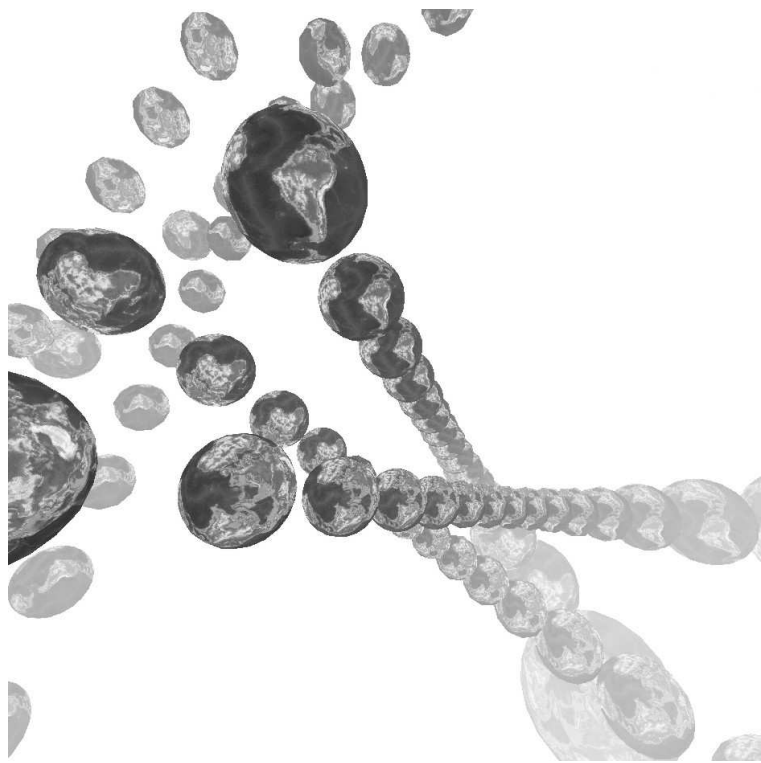, width=10cm}
 \caption{The cyclic factor $G'$ generates the images within
 each string, while the general factor $G$ takes one string to another.
 A $Z_3$ subgroup of $G$ permutes the three nearby strings.
 The axis of symmetry running along the
 center of the group of three strings is the common Clifford
 parallel, belonging to the family of left handed parallels of $G'$
 as well as the right handed parallels of the $Z_3$ subgroup of
 $G$.}
 \label{FigureStrings}
\end{center}
\end{figure}

Recall from Section \ref{SectionClassification} that at least one
factor in a double action manifold or linked action manifold must
be cyclic. Without loss of generality, assume the second factor
$G'$ is the cyclic group $Z_n$. Geometrically, this cyclic factor
creates a string of $n$ images of each cosmic source, arranged
along a great circle (see Figure~\ref{FigureStrings}) and related
by left handed Clifford translations.  The first factor $G$, which
may or may not be cyclic, creates parallel copies of the string of
$n$ images (see Figure~\ref{FigureStrings}), related by right
handed Clifford translations.

The elements of $G' = Z_n$ are all powers of a single generator,
and therefore translate along the same set of left handed Clifford
parallels.  The elements of $G$ may make use of several sets of
right handed Clifford parallels, one set for each maximal cyclic
subgroup of $G$.  Each such set of right handed Clifford parallels
has exactly two great circles in common with the left handed
Clifford parallels of $G'$ (recall Section~\ref{SectionClifford}
and Figure~\ref{FigureParallels}). One such {\it common Clifford
parallel} lies along the axis of symmetry of the three strings of
Earths in Figure~\ref{FigureStrings}: it belongs to the set of
left handed parallels that translate the Earths within each
string, and also belongs to the set of right handed parallels that
take one string of Earths to another.

\begin{figure}
\begin{center}
 \epsfig{file=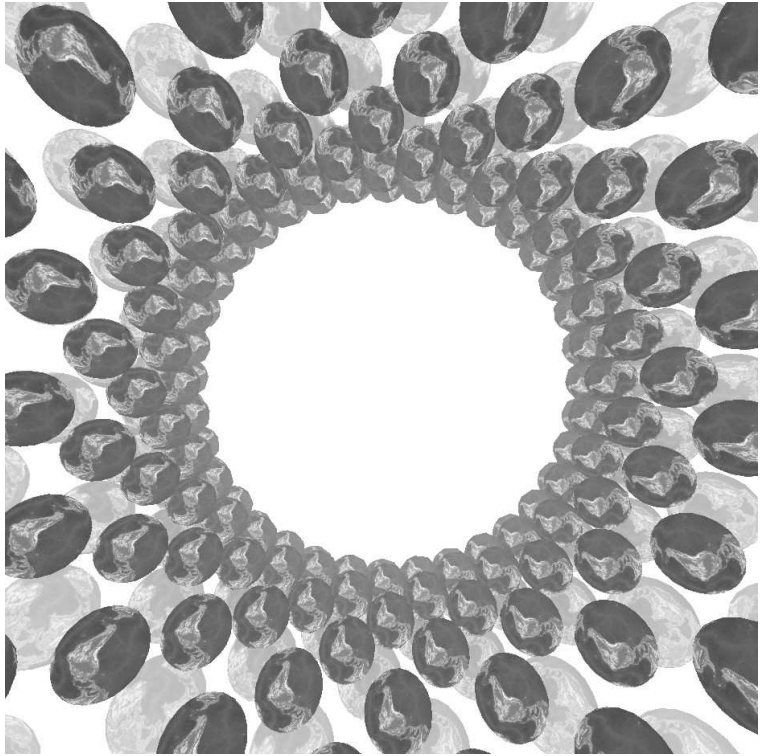, width=10cm}
 \caption{If both $G$ and $G'$ are cyclic of large order
 (here $G \approx Z_{17}$ and $G' \approx
 Z_{19}$), then many images of the observer crowd the invariant
 torus.  The axis of symmetry, orthogonal to the page, is the
 common Clifford parallel belonging to both $G$ and $G'$.  Note
 that the orientation of the images varies smoothly.}
\label{FigureCrowdedTorus}
\end{center}
\end{figure}

Both sets of Clifford translations preserve the {\it invariant
tori} centered around the common Clifford parallel (see
Figure~\ref{FigureParallels}). If both sets of Clifford
translations have large order, many images of the observer may
crowd the invariant torus (see Figure~\ref{FigureCrowdedTorus}).
In the exceptional case that the two sets of Clifford translations
have similar order, the closest distance between two images need
not be a pure left or right handed Clifford translation, but can
be some linear combination of the two.

Recall that every holonomy of a spherical 3-manifold can be
expressed as the composition of a left handed Clifford translation
and a right handed Clifford translation
(Section~\ref{SectionClassification}), and any left handed
Clifford translation shares a common axis with any right handed
one (see Figure~\ref{FigureParallels}). Thus every holonomy of a
spherical 3-manifold is a corkscrew motion along the common axis
of its left and right handed Clifford factors, and so the
situation described above is completely general. Nearby images may
depend on only one factor (as in Figure~\ref{FigureStrings}) or on
both factors (as in Figure~\ref{FigureCrowdedTorus})

If nearby images depend on only one factor
(Figure~\ref{FigureStrings}) then only the translates within a
single string produce matched circles. Searching for such matched
circles is easy:  they all come from Clifford translations, so
just as in Section~\ref{SubsectionSingleAction} the center of one
matched circle completely determines the center of its mate, and
the search space is either 4-dimensional (with no {\it a priori}
knowledge) or 3-dimensional (if $\chi_{\rm LSS}$ is known in
radians). If such circles are found, we detect the cyclic factor
$G' = Z_n$, but remain completely unaware of the other factor $G$.
Reversing this logic, if matched circles reveal a cyclic group in
the real universe, we cannot automatically conclude that the
universe is a lens space, but must consider the possibility of a
second factor.

\begin{figure}
\begin{center}
 \epsfig{file=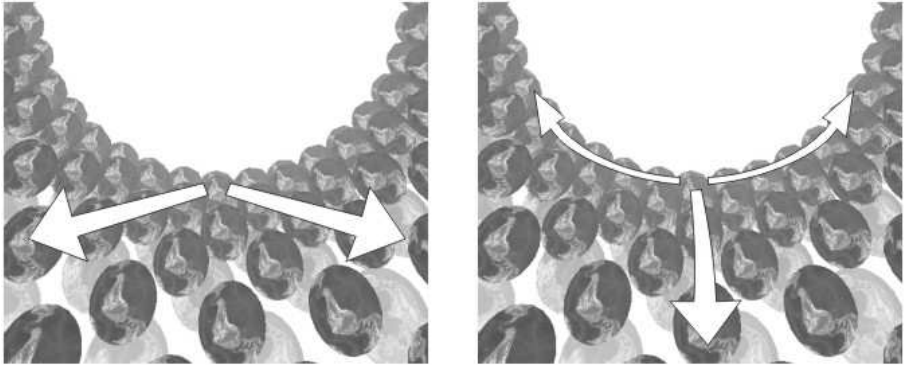, width=12cm}
 \caption{Translations along the
 Clifford parallels (left) keep matched circles directly opposite
 their mates while introducing a small twist.  Translations in the
 directions of maximal curvature (right) offset the matched
 circles' centers but introduce no twist.}
 \label{FigureSaddleNbr}
\end{center}
\end{figure}

If nearby images depend on both factors
(Figure~\ref{FigureCrowdedTorus}) then we must consider the
geometry of the invariant torus more carefully. The surface of the
invariant torus is locally saddle-shaped. Translations along the
left- or right-handed Clifford parallels of
Figure~\ref{FigureParallels} are translations along the horizontal
lines in Figure~\ref{FigureSaddleNbr}(left);  these translations
keep matching circles opposite each other in the sky, but
introduce a twist. Pure meridional or longitudinal rotations in
Figure~\ref{FigureParallels} are motions in the directions of
maximal curvature in Figure~\ref{FigureSaddleNbr}(right); they
displace matching circles so that their centers are no longer
exactly opposite, but introduce no twist.  Translations in generic
directions cause a small displacement of matching circles along
with a small twist. The magnitude of the displacement and twist
could be large if the Earth happens to lie near the common
Clifford parallel, but generically it is small.  Thus we must
search for circles in a full 6-parameter space, but we may assume
that the centers of the matching circles are almost opposite each
other and the twist angle is small.

\subsection{Matched circles search strategy}

The preceding paragraphs tell how to look efficiently for matching
circles in a nearly flat spherical universe:
\begin{itemize}
\item look for directly opposite circles to detect a single action
manifold or a single factor in a double action or linked action
manifold (Figure~\ref{FigureStrings})
\item or allow small deviations in circle position and twist to detect
a more general cyclic subgroup in a double or linked action
manifold (Figure~\ref{FigureCrowdedTorus}).
\end{itemize}
\vskip0.5cm

Let us now turn to the question of how much of the topology can be
detected using only nearby images of the observer. The cyclic
factor $G' = Z_n$ creates matching circles whenever $\Omega$ 
is greater than about $1 + {1}/{n^2}$ (see
Table~\ref{TableSingleActionCMB}). To understand the general
factor $G$, think of it as the (non disjoint) union of its maximal
cyclic subgroups.  Recall that the flow lines of each cyclic
subgroup $H \subset G$ share a unique pair of Clifford parallels
with the flow lines of $G'$ (Figure~\ref{FigureParallels}). If the
order of $H$ is small, then the parallel strings of images are
typically too far apart to create matching circles
(Figure~\ref{FigureStrings}), and $H$ remains undetectable. (An
exceptional case occurs when the observer happens to sit near the
common Clifford parallel, in which case the parallel strings of
images do get close enough to create matching circles.) If the
order of $H$ is large, then parallel strings of images may be
close enough to create matching circles
(Figure~\ref{FigureCrowdedTorus}). Consulting
Table~\ref{TableS3Subgroups}, let us consider each possibility for
the general factor $G$.
\begin{enumerate}
\item If $G$ is the binary tetrahedral group $T^*$, the binary
octahedral group $O^*$, or the binary icosahedral group $I^*$,
then the largest cyclic subgroup has order 6, 8, or 10,
respectively; however such a subgroup contains the antipodal map,
so it creates only 3, 4, or 5 strings of images, respectively. If
$\chi_{\rm LSS}$ is small (in radians), such parallel strings
create no matching circles, and $G$ remains undetectable. In the
exceptional case that the Earth lies near a common Clifford
parallel, we would detect a single cyclic subgroup of $G$ of order
at most 5.
\item If $G$ is a cyclic group $Z_k$, it will create
matched circles if and only if $k$ is large or the Earth happens
to lie near one of the pair of common Clifford parallels that $G$
shares with $G'$. In this case all images of the observer lie on a
single invariant torus, as shown in
Figure~\ref{FigureCrowdedTorus}.
\item If $G$ is a binary dihedral group $D^*_m$ we get all the
images on the invariant torus defined by its cyclic subgroup
$Z_{2m}$ as in the previous case
(Figure~\ref{FigureCrowdedTorus}), along with a second copy of the
invariant torus. Each of the many order 4 elements in $D^*_m$
interchanges the two invariant tori. In general the two invariant
tori are too far apart to create matched circles, so a $D^*_m
\times Z_n$ universe would in practice look like a $Z_{2m} \times
Z_n$ universe.  In the exceptional case that the two invariant
tori nearly coincide (which occurs only when the observer happens
to sit approximately midway between the two common Clifford
parallels), an image of the last scattering surface on one
invariant torus may intersect an image on the other;  it is
straightforward to see that the two matching circles would lie in
the same part of the sky and overlap each other, but such
non-generic behavior is unlikely to be observed in practice. In
practice one expects the generic case that matching circles lie
approximately (or exactly) opposite one other.
\end{enumerate}

\section{Cosmic Crystallography in a Nearly Flat Spherical Universe}
\label{SectionCrystallography}

The pair separation histogram (PSH) method for detecting cosmic
topology \cite{LLL} relies on the fact that in certain multiply
connected topologies, if the comoving distance from one source
(for example a galaxy cluster) to its nearest (resp. second
nearest, etc.) translated image is $\chi_{\rm gg}$, then the
distance from every source to its nearest (resp. second nearest,
etc.) translated image is also $\chi_{\rm gg}$. Thus to seek
topology, one measures the distances between all pairs of sources
(without worrying about which are topological images of one
another) and checks whether any distances $\chi_{\rm gg}$ occur
extraordinarily often. Crystallographic methods probe topology on
the scale of the catalog's depth $\chi_{\rm catalog} = \chi(z)$
for $z \sim 1-3$, so $\chi_{\rm LSS}$ has to be replaced by
$\chi_{\rm catalog}$ in the previous discussion.

Among the hyperbolic topologies, the PSH method never
works~\cite{lehoucq99}, because the distance from a source to its
nearest translated image always depends on the source's location
in the manifold. Among the flat topologies, the PSH method works
for the 3-torus, because the distance from a source to its nearest
image does not depend on its location, but for all other flat
topologies the PSH method detects only a subgroup of the full
holonomy group, namely the subgroup of pure translations.  Among
the spherical topologies, the PSH method detects precisely the
Clifford translations (despite all observational
uncertainties~\cite{lehoucq00}).

\begin{table}
\begin{center}
\begin{tabular}{|l|c|c|}
\hline
Group & Min. translation distance & Sample min. $\Omega$ \\
\hline
Binary icosahedral $I^*$   & $2\pi/10$ & 1.048 \\
Binary octahedral  $O^*$   & $2\pi/8 $ & 1.073 \\
Binary tetrahedral $T^*$   & $2\pi/6 $ & 1.125 \\
Binary dihedral    $D^*_m$ & $2\pi/2m$ & $\sim (1 + {5}/{(2m)^2})$ \\
Cyclic             $Z_n$   & $2\pi/n$  & $\sim (1 + {5}/{n^2})$ \\
\hline
\end{tabular}
\end{center}
 \caption{Crystallographic methods probe topology only to a
 distance of $\chi(z)$ for $z \sim 3$, in which case
 the minimum values of $\Omega$ given in
 Table~\ref{TableSingleActionCMB} for circle matching
 at $\chi(z_{\rm LSS})$ must be replaced by the values shown
 here for crystallographic studies at $\chi(3)$.
 That is, a $z \leq 3$ catalog may contain repeated images
 of galaxy clusters in the given topology if and only if
 $\Omega$ exceeds the value given.
 As in Table~\ref{TableSingleActionCMB}
 the sample values in the third column fix $\Omega_m$ at 0.35 and
 vary $\Omega_{\Lambda}$.}
 \label{TableSingleActionPSH}
\end{table}

In a single action manifold
(Subsection~\ref{SubsectionSingleAction}) all holonomies are
Clifford translations, and thus the topology is potentially
detectable if and only if $\Omega$ exceeds the minimum
value as illustrated in Table~\ref{TableSingleActionPSH}.  In the
case of the binary tetrahedral, binary octahedral, and binary
icosahedral groups we would detect equal length translations in
several different directions, and could infer the topology
exactly.  In the case of a binary dihedral group (assuming
$\Omega \approx 1$) we would see only the cyclic
factor, so in practice we couldn't distinguish the binary dihedral
group from a cyclic group.

In a double action manifold (Subsection
\ref{SubsectionDoubleAction}) the PSH could detect the factors $G$
and $G'$ individually, subject to the conditions explained in the
preceding paragraph, because each is a group of Clifford
translations, but it could not detect any nontrivial composite
elements $(g,g') \in G \times G'$.  Fortunately our nearest
translates will typically be Clifford translates (and thus
potentially detectable) even though the vast majority of other
translates are not (cf. Figure~\ref{FigureStrings}).  Only in
exceptional cases (namely when $G$ and $G'$ are cyclic of
comparable order, or the observer is at a non-generic location)
might our nearest translates fail to be Clifford translates.

In a linked action manifold (Subsection
\ref{SubsectionLinkedAction}) only a subgroup of each factor $G$
and $G'$ is available, so it is harder to detect a linked action
manifold than a double action one.  In some cases, such as the
lens space $L(5,2)$, the available subgroups are trivial and the
PSH cannot detect the topology at all.  Fortunately if one
excludes the lens spaces then the remaining linked action
manifolds are index 2 or index 3 subgroups of the corresponding
double action manifolds, the picture is qualitatively the same as
in Figure \ref{FigureStrings}, and again the nearest translates
are Clifford translates, which the PSH can detect.

\section{Conclusion}

In this article, we have studied the detectability of topology
when the universe is spherical but nearly flat, both by the
CMB-based circle-in-the-sky method and the large scale
structure-based crystallographic method.

Spherical topologies turn out to be the most readily detectable.
After recalling their geometrical properties, we first showed that
the circles-in-the-sky method typically detects only a cyclic
subgroup of the holonomy group, so the universe ``looks like a
lens space'' {\it no matter what its true topology is}. If it
looks like a globally homogeneous lens space, then the
circles-in-the-sky search reduces from a 6-parameter search space
to a 3-parameter space  or a 4-parameter space depending on
whether the radius of the last scattering surface is known or not.
If the universe looks like a non globally homogeneous lens space
then the matching circles lie in a 6-parameter space, but with
strong constraints on their values, so that the search is still
much faster than in the general case.  We also showed that
crystallographic methods are well suited to detect the topology of
a nearly flat spherical universe thanks to the abundance of
Clifford translations.

We hope our analyses and results will serve as preparation for
intelligently searching the upcoming MAP satellite data for a
topological signal.

\ack{We thank Simon Prunet for discussions. JRW thanks the
MacArthur Foundation for its support.} \vskip0.5cm

\end{document}